\documentclass[aps,twocolumn,preprintnumbers,superscriptaddress,showpacs,nofootinbib,amsmath, amssymb]{revtex4-2}

\usepackage{graphicx}
\usepackage{subfigure}
\usepackage{hyperref}
\usepackage{color}
\usepackage[table]{xcolor}
\usepackage{boldline,multirow}
\usepackage{booktabs}
\usepackage{pifont}
\usepackage{mathrsfs}
\usepackage{soul}
\usepackage[utf8]{inputenc}
\usepackage{multirow}

\usepackage{color}
\definecolor{rosso}{cmyk}{0,1,1,0.4}
\definecolor{rossos}{cmyk}{0,1,1,0.55}
\definecolor{rossoc}{cmyk}{0,1,1,0.2}
\definecolor{blu}{cmyk}{1,1,0,0.3}
\definecolor{blus}{cmyk}{1,1,0,0.6}
\definecolor{bluc}{cmyk}{1,1,0,0.1}
\definecolor{verde}{cmyk}{0.92,0,0.59,0.25}
\definecolor{verdec}{cmyk}{0.92,0,0.59,0.15}
\definecolor{verdes}{cmyk}{0.92,0,0.59,0.4}
\definecolor{bviolet}{rgb}{0.54, 0.17, 0.89}
\definecolor{myred}{rgb}{0.545,0.004,0}
\hypersetup{colorlinks,bookmarksopen,bookmarksnumbered,citecolor=myred,
linkcolor=myred,pdfstartview=FitH,urlcolor=myred}

\newcommand{\beq}{\begin{equation}} 
\newcommand{\eeq}{\end{equation}}
\newcommand{\bea}{\begin{eqnarray}}  
\newcommand{\eea}{\end{eqnarray}}
\newcommand{\beastar}{\begin{eqnarray*}}  
\newcommand{\eeastar}{\end{eqnarray*}}
\newcommand{\nnl}{\nonumber \\}

\definecolor{Gray}{gray}{0.9}

\begin{document}

\preprint{DESY 20-190}
\preprint{HU-EP-20/30}
\preprint{TTP20-037}
\preprint{P3H-20-064}
\preprint{UCI-TR 2020-18}

\title{Lessons from the \texorpdfstring{\boldmath$B^{0,+}\to K^{*0,+}\mu^+\mu^-$}{B to Kst mu+ mu-} angular analyses}

\author{Marco Ciuchini}
\email[]{marco.ciuchini@roma3.infn.it}
\affiliation{INFN Sezione di Roma Tre,
Via della Vasca Navale 84, I-00146 Rome, Italy}

\author{Marco Fedele}
\email[]{marco.fedele@kit.edu}
\affiliation{Institut f\"ur Theoretische Teilchenphysik, Karlsruhe Institute of Technology, D-76131 Karlsruhe, Germany}

\author{Enrico Franco}
\email[]{enrico.franco@roma1.infn.it}
\affiliation{INFN Sezione di Roma, Piazzale Aldo Moro 2, I-00185 Rome, Italy}

\author{Ayan Paul}
\email[]{ayan.paul@desy.de}
\affiliation{DESY, Notkestra{\ss}e 85, D-22607 Hamburg, Germany}
\affiliation{Institut f\"ur Physik, Humboldt-Universit\"at zu Berlin, D-12489 Berlin, Germany}

\author{Luca Silvestrini}
\email[]{luca.silvestrini@roma1.infn.it}
\affiliation{INFN Sezione di Roma, Piazzale Aldo Moro 2, I-00185 Rome, Italy}

\author{Mauro Valli}
\email[]{mvalli@uci.edu}
\affiliation{Department of Physics and Astronomy, University of California, Irvine,	California 92697, USA}

\begin{abstract}
We perform an analysis within the Standard Model of $B^{0,+} \to K^{*0,+} \mu^+ \mu^-$  decays in light of the recent measurements from the LHCb experiment, showing that new data strengthen the need for sizable hadronic contributions and correlations among them. 
We then extend our analysis to New Physics via the Standard Model Effective Theory, and carry out a state-of-the-art fit of available $b \to s \ell^+ \ell^-$ data, including possible hadronic contributions. We find the case of a fully left-handed operator standing out as the simplest scenario with a significance of almost $6\sigma$.
\end{abstract}

\maketitle

After the observation of the Higgs boson \cite{Aad:2012tfa,Chatrchyan:2012ufa}, indirect searches for physics beyond the Standard Model (SM) are playing an increasingly important role in the program of the Large Hadron Collider (LHC), as the recorded luminosity increases. 
In addition to precision electroweak and Higgs physics, LHC is also providing a huge amount of high-precision data in the flavour sector, in particular on rare and CP-violating decays of heavy mesons.
In this context, $b \to s \ell^+ \ell^-$ transitions have recently been under the spotlight, not only because of their potential sensitivity to New Physics (NP)~\cite{Hiller:2014yaa,Hiller:2014ula,Hurth:2014vma,Descotes-Genon:2015uva}, 
but also because of the current experimental hints of deviations from the SM, see, e.g.,~\cite{DAmico:2017mtc,Geng:2017svp,Capdevila:2017bsm,Ciuchini:2017mik,Hiller:2017bzc,Alok:2017sui,Ciuchini:2019usw,Aebischer:2019mlg,Kowalska:2019ley,Alguero:2019ptt,Datta:2019zca,Arbey:2019duh}. 
As any other indirect search for NP, the quest for NP in $b \to s \ell^+ \ell^-$ decays requires not only high experimental precision, but also a robust estimate of theoretical uncertainties in the SM prediction.
From this point of view, the set of experimental results which hint at NP in $b \to s \ell^+ \ell^-$ transitions can be divided in two broad classes. The first contains
ratios of decay Branching Ratios (BRs) for different leptons in the final state; the second contains absolute BRs and angular distributions.
The former is particularly clean from the theoretical point of view~\cite{Hiller:2003js,Bordone:2016gaq,Isidori:2020acz}, but experimentally challenging,\footnote{Ratios of angular observables as the ones proposed in \cite{Capdevila:2016ivx,Serra:2016ivr,Alguero:2019pjc} and measured by Belle in \cite{Wehle:2016yoi} may also be considered in this category.} while the latter is also subject to sizable theoretical uncertainties ~\cite{Jager:2012uw,Jager:2014rwa}.  
Indeed, while the calculation of decay amplitudes for exclusive $b \to s \ell^+ \ell^-$ transitions is well-defined in the infinite $b$ and $c$ mass limit \cite{hep-ph/9905312,Beneke:2000ry,Beneke:2004dp}, and 
while in the same limit the uncertainty from decay form factors can be eliminated by taking suitable ratios of observables~\cite{Descotes-Genon:2013vna,Descotes-Genon:2014uoa}, in the real world amplitude calculations must 
cope with power corrections, which can be sizable or even dominant in several kinematic regions~\cite{Ciuchini:2015qxb,Ciuchini:2016weo,Arbey:2018ics,Ciuchini:2018anp,Hurth:2020rzx}. For example, the Operator Product Expansion is known to fail altogether for resonant $B \to K^{(*)} J/\psi \to K^{(*)} \mu^+ \mu^-$ transitions \cite{Beneke:2009az}, and its accuracy is questionable close to the $c \bar c$ threshold. 
For this reason, estimating corrections to QCD factorization in the low dilepton invariant mass (low-$q^2$) region of $B \to K^{(*)} \ell^+ \ell^-$ decay amplitudes is a crucial step towards a reliable assessment of possible deviations from SM predictions in these decay channels. 
Unfortunately, first-principle calculations of these power corrections are not currently available, and a theoretical breakthrough would be needed to perform such calculations, see, e.g., the discussion in~\cite{Jager:2014rwa,Bobeth:2017vxj,Melikhov:2019esw}. Waiting for this breakthrough, the only reliable option is to use data-driven methods to account for the theoretical uncertainties and to quantify possible deviations from the SM. 
Obviously, data-driven methods are (much) less NP sensitive than (bold) theoretical assumptions, but as more and more data become available the road to a robust test of 
the SM becomes viable. 
In this context, the very recent angular analysis of the $B^+ \to K^{*+} \mu^+ \mu^-$ decay~\cite{LHCb_new_angular}, together with the recent update on the $B^0 \to K^{*0} \mu^+ \mu^-$ one~\cite{Aaij:2020nrf},  represents a milestone in the effort to disentangle possible NP contributions from
long-distance QCD effects. 
In this \textit{Letter}, we exploit these recent data to perform a detailed study of QCD pollution in angular observables, and to assess the
compatibility of $B^{0,+} \to K^{*0,+} \mu^+ \mu^-$ with the SM. We then combine angular observables with Lepton Flavour Universality (LFU) violating ones to provide the best
estimate of possible NP contributions to $b \to s \ell^+ \ell^-$ transitions. 
The lesson we learn from the present analysis is twofold: 
\textit{i)} Within the SM, experimental data on angular analyses can be reproduced with
sizable hadronic contributions, including a possible contribution that mimics NP effects; 
\textit{ii)} In the Standard Model Effective Theory (SMEFT) \cite{Buchmuller:1985jz,Grzadkowski:2010es}, the significance of NP from the global $b \to s \ell^+ \ell^-$ analysis increases with the inclusion of new data, reaching a maximum of almost $6\sigma$ for the 
simple scenario of a non-vanishing $C^{LQ}_{2223}$, always taking into account hadronic effects (see eq.~\eqref{eq:SMEFT_op_tree} below for the definition).
\noindent All details of our treatment of hadronic uncertainties and of our Bayesian analysis technique can be found in refs. \cite{Ciuchini:2015qxb,Ciuchini:2018anp,Ciuchini:2019usw}; here we limit ourselves to a concise review of the necessary ingredients. The main contributions to the $B \to K^{(*)}\ell^+\ell^-$ decay amplitudes come from the following operators:
\bea
\label{eq:Q7}
Q_{7\gamma} &=& \sqrt{\frac{\alpha_{e}}{64 \pi^3}} m_b \bar{s}_L\sigma_{\mu\nu}F^{\mu\nu}b_R\,, \\
\label{eq:Q9}
Q_{9V,\ell} &=& \frac{\alpha_{e}}{4\pi}(\bar{s}_L\gamma_{\mu}b_L)(\bar{\ell}\gamma^{\mu}\ell)\,, \\
\label{eq:Q10}
Q_{10A,\ell} &=& \frac{\alpha_{e}}{4\pi}(\bar{s}_L\gamma_{\mu}b_L)(\bar{\ell}\gamma^{\mu}\gamma^5\ell) \,,\\
 \label{eq:quark_O}
Q^c_2 &=& (\bar{s}_L\gamma_{\mu} c_L)(\bar{c}_L\gamma^{\mu}b_L)\,.
\eea
Following~\cite{Jager:2012uw,Gratrex:2015hna}, SM decay amplitudes can be conveniently decomposed in the helicity basis:
\bea \label{eq:helamp}
H_V^{\lambda} &\propto& \left\{C_9^{\rm SM}\widetilde{V}_{L\lambda} + \frac{m_B^2}{q^2} \left[\frac{2m_b}{m_B}C_7^{\rm SM}\widetilde{T}_{L\lambda}  - 16\pi^2h_{\lambda} \right]\right\}\,,\nonumber\\
H_A^{\lambda} &\propto& C_{10}^{\rm SM}\widetilde{V}_{L\lambda} \ , \
H_P \propto \frac{m_{\ell} \, m_{b}}{q^2} \,  C_{\rm 10}^{\rm SM} \left( \widetilde{S}_{L} - \frac{m_s}{m_b}\widetilde{S}_{R} \right) \label{Hp}
\eea
with $\lambda=0,\pm$ and $C_{7,9,10}^{\rm SM}$ the SM Wilson coefficients of the operators in  eqs.~\eqref{eq:Q7}-\eqref{eq:Q10}, normalized as
in ref.~\cite{Ciuchini:2019usw}. 
The factorizable part of the amplitudes corresponds to seven independent form 
factors, $\widetilde{V}_{0,\pm}$, $\widetilde{T}_{0,\pm}$ and $\widetilde{S}$, smooth functions of $q^{2}$~\cite{Straub:2015ica,Gubernari:2018wyi}. At first order in $\alpha_{e}$, non-local effects arise from the insertion of the operator in eq.~\eqref{eq:quark_O} yielding non-factorizable power corrections in $H_V^{\lambda}$ via the hadronic correlator $h_\lambda(q^2)$~\cite{Jager:2014rwa,Ciuchini:2015qxb,Chobanova:2017ghn}, receiving the main contribution from the time-ordered product of:
\beq \label{eq:hlambda}
\frac{\epsilon^*_\mu(\lambda)}{m_B^2} \int d^4x\ e^{iqx} \langle \bar K^* \vert \mathcal{T}\{\bar{c}(x)\gamma^{\mu} c(x) 
Q^{c}_{2} (0)\} \vert \bar B \rangle \,.
\eeq
Within different setups and assumptions, most recent attempts to estimate the charm-loop contribution of eq.~\eqref{eq:hlambda} ~\cite{Blake:2017fyh,Bobeth:2017vxj,Chrzaszcz:2018yza} find agreement with the outcome of the light-cone sum-rule computation in~\cite{Khodjamirian:2010vf}. However, a reliable estimate of non-factorizable effects encoded in $h_{0,\pm}(q^2)$ remains theoretically challenging in the full kinematic region of interest. In this work, we adopt a data-driven method based on the following general parameterization of the hadronic contributions:
\begin{eqnarray} 
\label{eq:hv}
H_V^{-} \propto  
 & \frac{m_B^2}{q^2}&  \bigg[ \frac{2m_b}{m_B}\left(C_7^{\rm SM} + h_-^{(0)} \right) \widetilde T_{L -}  
  -  16\pi^2 h_-^{(2)}\, q^4 \bigg] \nnl 
  & + & \left(C_9^{\rm SM} + h_-^{(1)}\right)\widetilde V_{L -}\,, \nnl  
H_V^{+} \propto  
 & \frac{m_B^2}{q^2}&  \bigg[ \frac{2m_b}{m_B}\left(C_7^{\rm SM} + h_-^{(0)} \right) \widetilde T_{L +}  
  -  16\pi^2 \Big(h_{+}^{(0)}  \nnl 
  & + & h_{+}^{(1)}\, q^2 +  h_{+}^{(2)}\, q^4\Big) \bigg] + \left(C_9^{\rm SM} + h_-^{(1)}\right)\widetilde V_{L +}\,, \nnl 
H_V^{0} \propto  
 & \frac{m_B^2}{q^2}&  \bigg[ \frac{2m_b}{m_B}\left(C_7^{\rm SM} + h_-^{(0)} \right) \widetilde T_{L 0}  
  -  16\pi^2 \sqrt{q^2} \Big(h_{0}^{(0)}  \nnl 
  & + & h_{0}^{(1)}\, q^2 \Big) \bigg] + \left(C_9^{\rm SM} + h_-^{(1)}\right)\widetilde V_{L 0}\,.
\end{eqnarray}
It is evident from eq.~\eqref{eq:hv} that the coefficients $h_-^{(0)}$ and $h_-^{(1)}$ can mimic LFU effects of NP, contributing to $C_7$ and $C_9$ respectively. Consequently, the extraction of NP contributions to $C_{7,9}$ from angular observables crucially depends on the theoretical assumption on the size of $h_-^{(0,1)}$. 
However, precise experimental data can in principle lead to the determination of all $h$'s, improving our knowledge of hadronic contributions and strengthening or weakening our confidence on the estimates of refs.~\cite{Khodjamirian:2010vf,Blake:2017fyh,Bobeth:2017vxj,Chrzaszcz:2018yza}. In this context, it is very interesting to quantify the impact of the new data on the determination of the $h$'s. Using the \texttt{HEPfit} code \cite{deBlas:2019okz,HEPfit}, we compare the results of a SM fit to the data in refs.~\cite{Aaij:2012ita,Aaij:2015oid,Aaij:2016flj,Wehle:2016yoi,Aaboud:2018krd,Khachatryan:2015isa,Sirunyan:2017dhj,Aaij:2015dea,Aaij:2015esa,Aaij:2017vad,Chatrchyan:2013bka,Aaboud:2018mst,Aaij:2014pli,Aaij:2017vbb,Abdesselam:2019wac,Aaij:2019wad,Aaij:2020nrf,LHCb_new_angular} with the one omitting the most recent data in refs.~\cite{Aaij:2020nrf,LHCb_new_angular}.

\begin{figure}[!t!]
\includegraphics[width=\columnwidth]{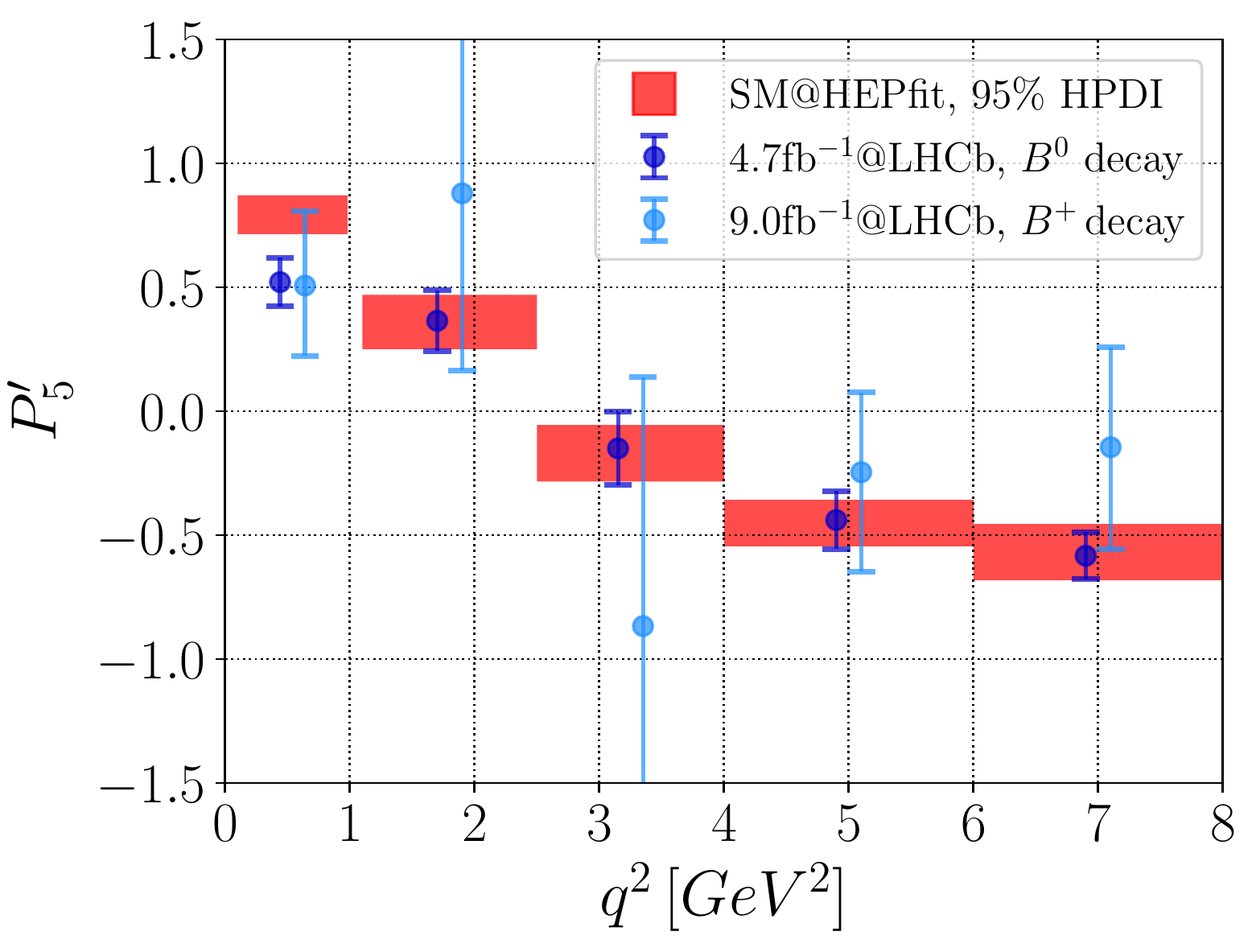}
\caption{\it Result from a fit in the SM to the up-to-date experimental $b \to s \ell^{+} \ell^{-}$ data at low $q^2$ for the binned angular observable $P^{'}_{5}$~\cite{DescotesGenon:2012zf}. We show the obtained 95\% highest probability density interval (HPDI) adopting the parameterization in eq.~\eqref{eq:hlambda}, together with the most recent measurements from the LHCb angular analyses in~\cite{Aaij:2020nrf,LHCb_new_angular}. Quark-spectator effects distinguishing the outcome for $B^{0,+}$ decays are at the percent level..}
\label{fig:P5p}
\end{figure}

\begin{figure*}[!ht!]
\includegraphics[width=\textwidth]{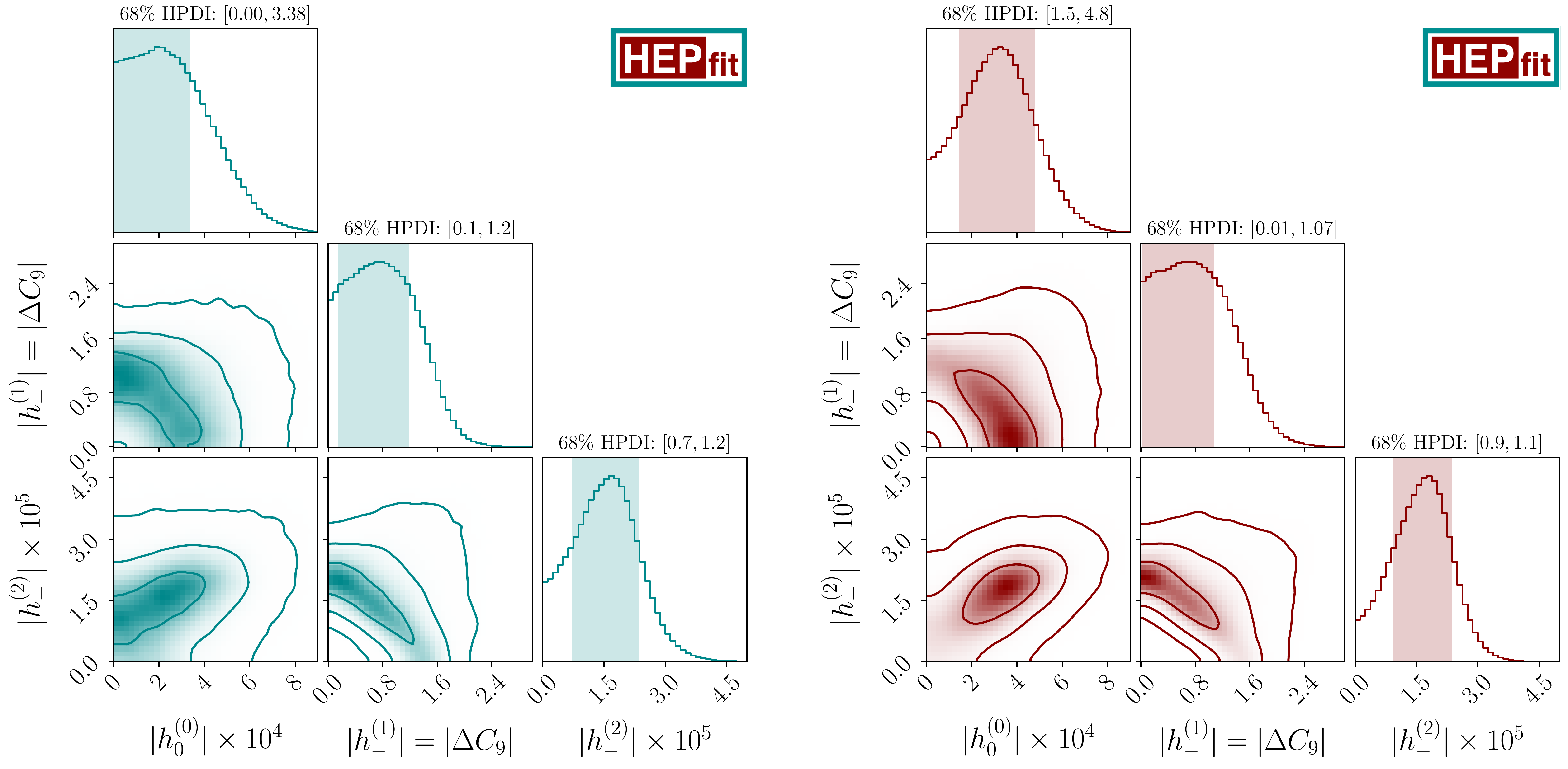}
\caption{\it Inference of hadronic contributions from a fit in the SM to the available experimental $b \to s \ell^{+} \ell^{-}$ dataset at low $q^2$, adopting the parameterization in eq.~\eqref{eq:hlambda}, omitting (left panel in green) or using (right panel in red) new data from refs.~\cite{Aaij:2020nrf,LHCb_new_angular}. 
Contours correspond to smallest regions of $68\%$, $95\%$, $99.7\%$ probability. For marginalized one-dimensional posterior distributions the $68\%$ highest probability density interval (HPDI) is explicitly reported, highlighted by vertical bands.}
\label{fig:SM}
\end{figure*}

Our main results in the SM are presented in Figs.~\ref{fig:P5p}-\ref{fig:SM}, where the impact of the new data on the determination of the hadronic contributions (including $h_-^{(1)} \equiv \Delta C_9$) can be clearly seen. In particular, in Fig.~\ref{fig:P5p} we show how the latest experimental information on $P_{5}'$, see ref.~\cite{DescotesGenon:2012zf}, can be accommodated in the SM once sizable hadronic effects as the ones obtained for $B^{0,+} \to K^{*0,+} \ell^{+} \ell^{-}$ in Fig.~\ref{fig:SM} are taken into account.
In the left panel of Fig.~\ref{fig:SM} we present an update of our analysis of ref.~\cite{Ciuchini:2019usw}, studying all available $b \to s \ell^{+} \ell^{-} $ data at low $q^2$ previous to the LHCb measurements in~\cite{Aaij:2020nrf,LHCb_new_angular}.
In the right panel we then show the impact of the latter set: \textit{i)} the evidence of a non-vanishing combination of $\vert h_-^{(1)} \vert$ and $\vert h_-^{(2)}\vert$ is strengthened, with a slight ($<2\sigma$) preference for a non-vanishing $\vert h_-^{(2)}\vert$; \textit{ii)} a new evidence of a non-vanishing combination of $\vert h_-^{(1)}\vert $ and $\vert h_0^{(0)}\vert$ emerges, with a slight ($<2\sigma$) preference for a non-vanishing $\vert h_0^{(0)}\vert$. Thus, new data globally strengthen the evidence of non-vanishing $h$'s, introducing a slight preference for purely hadronic contributions.

Generalizing our analysis to the SMEFT, we consider the following additional operators:
\bea \label{eq:SMEFT_op_tree}
O^{LQ^{(1)}}_{2223} &=& (\bar{L}_2\gamma_\mu L_2)(\bar{Q}_2\gamma^\mu Q_3)\,, \nonumber \\
O^{LQ^{(3)}}_{2223} &=& (\bar{L}_2\gamma_\mu \tau^{A} L_2)(\bar{Q}_2\gamma^\mu\tau^{A} Q_3)\,,\nonumber \\
O^{Qe}_{2322} &=& (\bar{Q}_{2}\gamma_\mu Q_{3})(\bar{e}_{2} \gamma^\mu e_{2})\,,\nonumber \\
O^{Ld}_{2223} &=& (\bar{L}_2\gamma_\mu L_2)(\bar{d}_2\gamma^\mu d_3)\,,\nonumber \\
O^{ed}_{2223} &=& (\bar{e}_2\gamma_\mu e_2)(\bar{d}_2\gamma^\mu d_3)\,,
\eea
where $\tau^{A=1,2,3}$ are Pauli matrices (a sum over $A$ in the equations above is understood), weak doublets are in upper case and $SU(2)_{L}$ singlets are in lower case, and flavour indices are defined in the basis of diagonal down-type quark Yukawa couplings. Since in our analysis operators $O^{LQ^{(1,3)}}_{2223}$ always enter as a sum, we collectively denote their Wilson coefficient as $C^{LQ}_{2223}$. We normalize SMEFT Wilson coefficients to a NP scale $\Lambda = 30$ TeV. With this normalization, after electroweak symmetry breaking $C_{9}$ receives contributions from $\mathcal{N}_\Lambda (C^{LQ}_{2223}+C^{Qe}_{2322})$, $C_{10}$ from $\mathcal{N}_\Lambda (-C^{LQ}_{2223}+C^{Qe}_{2322})$ and the chirality-flipped operators $C_{9}^\prime$ from $\mathcal{N}_\Lambda (C^{ed}_{2223}+C^{Ld}_{2223})$, $C_{10}^\prime$ from $\mathcal{N}_\Lambda (C^{ed}_{2223}-C^{Ld}_{2223})$, with $\vert\mathcal{N}_\Lambda \vert \simeq 0.7$. To quantitatively compare different NP scenarios, where different sets of SMEFT Wilson coefficients are allowed to float, to the SM, we compute the \textit{Information Criterion} (IC) \cite{IC}: 
\begin{equation}\label{eq:IC}
   IC \equiv -2 \overline{\log \mathcal{L}} \, + \, 4 \sigma^{2}_{\log \mathcal{L}} \,,
\end{equation}
where the first and second terms respectively represent mean and variance of the loglikelihood posterior distribution.
Model selection between two scenarios proceeds according to the smallest IC value reported and the extent to which a model should be preferred over another one follows the canonical scale of evidence of ref.~\cite{BayesFactors}, related in this context to (positive) IC differences. For convenience we always report $\Delta IC \equiv IC_{\textrm{SM}} - IC_{\textrm{NP}}$.

\begin{table}[!t!]
\centering
\renewcommand{\arraystretch}{1.5}
{\footnotesize
\begin{tabular}{|c|cc|}
\hline
NP scenario & mean(std) & $\Delta IC$ \\
\hline 
\multirow{2}{*}{A:  $ \ C^{LQ}_{2223} \ $ \ \  \ \ \  \ \ \ }
& 0.77(13) & 29 \\
&\cellcolor{Gray} 0.92(12) & \cellcolor{Gray} 58 \\
\hline
\multirow{2}{*}{B: $ \{C^{LQ}_{2223},C^{Qe}_{2322}\} $}
& \{0.80(18), 0.05(30)\} & 26 \\
&\cellcolor{Gray} \{1.03(12), 0.71(13)\} &\cellcolor{Gray} 81 \\
\hline
C: $\{C^{LQ}_{2223},C^{Qe}_{2322},$
& \{1.11(23), 0.49(36), -0.42(23), -0.28(43)\} & 26 \\
$\ \ \ \ C^{Ld}_{2223},C^{ed}_{2223}\}$
&\cellcolor{Gray} \{1.10(12), 0.83(15), -0.33(19), 0.04(37)\} &\cellcolor{Gray} 89 \\
\hline
\end{tabular}
}
\caption{ \it Mean and standard deviation (std) of the posterior distribution of the SMEFT Wilson coefficients from a fit to the full set of most recent $b \to s \ell^{+} \ell^{-}$ data at low $q^2$ in the NP scenarios A,B,C along with $\Delta IC \equiv IC_{\textrm{SM}} - IC_{\textrm{NP}}$. Results in white lines are obtained allowing for hadronic contributions as in the parameterization in eq.~\eqref{eq:hlambda}, while results in gray lines are obtained using the $q^2$ extrapolation of the QCD sum-rule estimates of~\cite{Khodjamirian:2010vf}. 
\label{tab:WC_SMEFT}}
\end{table}
\begin{figure*}[!t!]
\includegraphics[width=\textwidth]{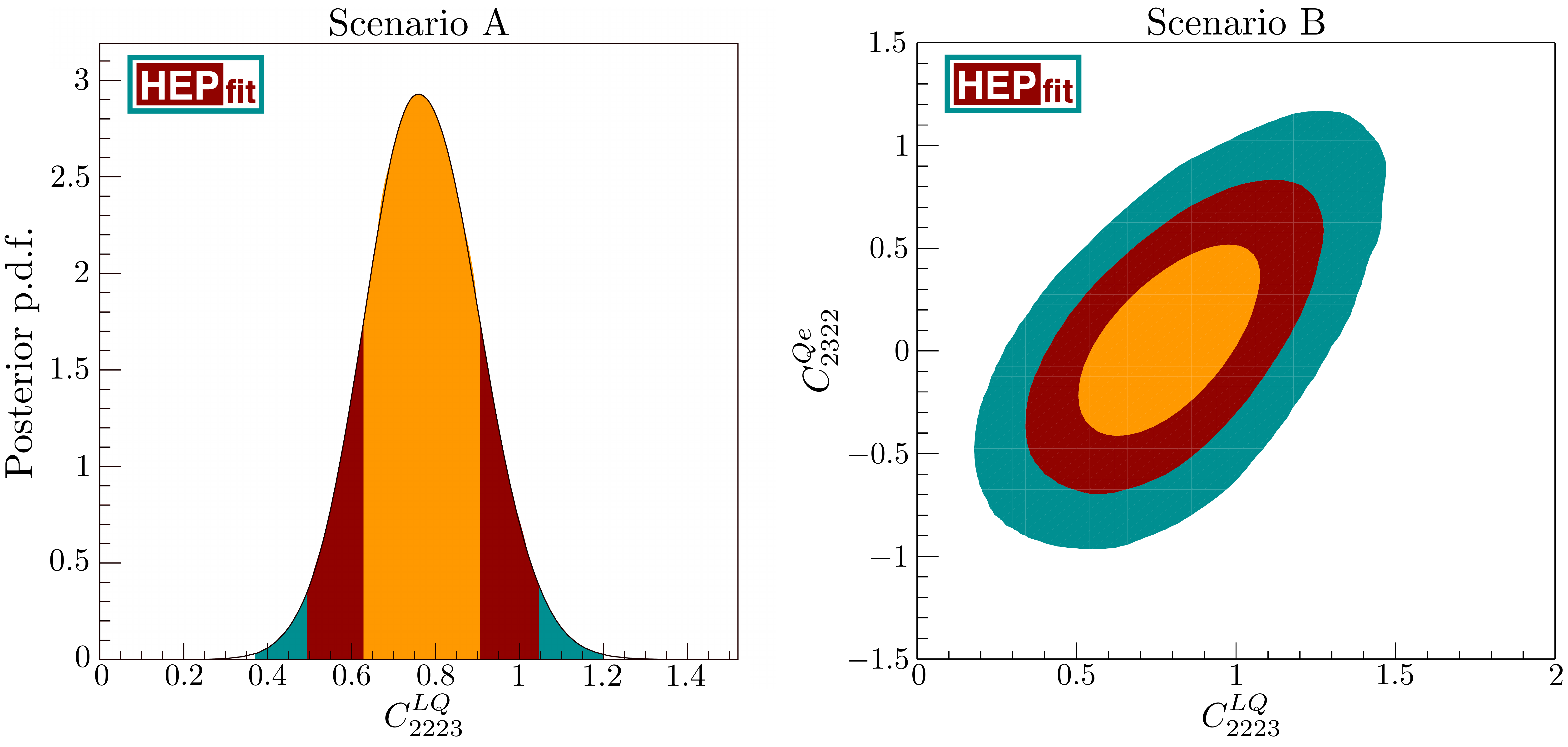}
\caption{\it \underline{Left panel}: Posterior probability density function (p.d.f.) for the NP coefficient $C^{LQ}_{2223}$ in scenario A. \underline{Right panel}: Joint posterior p.d.f for $C^{LQ}_{2223}$ and $C^{Qe}_{2322}$ in scenario B. We show $68\%$, $95\%$ and $99.7\%$ probability regions in orange, red and green respectively. All results are obtained using the parameterization of hadronic contributions in eq.~\eqref{eq:hlambda}.}
\label{fig:CLQ}
\end{figure*}

In the simplest NP scenario considered (scenario A), we just allow for NP contributions to appear in $C^{LQ}_{2223}$, corresponding to $\Delta C_{9,\mu} = - \Delta C_{10,\mu}$. We then generalize to the case of non-vanishing $C^{LQ}_{2223}$ and $C^{Qe}_{2322}$ (scenario B), which allows for independent NP contributions to $C_{9,\mu}$ and $C_{10,\mu}$. 
Finally, we also switch on $C^{ed}_{2223}$ and $C^{Ld}_{2223}$, thus allowing for NP to modify independently also the chirality-flipped operators $C_{9,\mu}^{\prime}$ and $C_{10,\mu}^{\prime}$ (scenario C). The results of our fit in the three scenarios described above are summarized in Table \ref{tab:WC_SMEFT} and Figure \ref{fig:CLQ}. Our main conclusion is that the preferred scenario is the simplest one, namely a NP contribution to $C^{LQ}_{2223}$, or equivalently $\Delta C_{9,\mu} = - \Delta C_{10,\mu}$, leading to $\Delta IC = 29$. The fitted value of $C^{LQ}_{2223}=0.77 \pm 0.13$ corresponds to $\Delta C_{9,\mu} = - \Delta C_{10,\mu} = -0.54 \pm 0.09$ for a NP scale $\Lambda$ of 30 TeV, deviating from the SM with a significance of $\sim6\sigma$. Scenarios B and C, in spite of the increase in model complexity, do not produce a sizable improvement in the fit. 

The conclusion would be very different if a less conservative approach to hadronic uncertainties was taken, using QCD sum-rule estimates of the hadronic contributions and extrapolating them to the whole kinematic range up to the largest $q^2$ bin in Fig. \ref{fig:SM}. Then, the simplest scenario would not lead to an optimal description of experimental data, and additional operators would be needed.
From the grey lines in Table \ref{tab:WC_SMEFT}, the four-operator scenario including chirality-flipped operators achieves the best result, reproducing a NP pattern similar to the one with simultaneously non-vanishing $(C_{9,\mu},C_{10,\mu}^\prime)$ highlighted, e.g., in~\cite{Ciuchini:2019usw,Alguero:2019ptt}. We stress again that a conservative treatment of hadronic uncertainties is therefore crucial to obtain an unbiased picture of the kind of NP that may lie behind these intriguing experimental results. 

Future updates of the present fit with forthcoming experimental data from LHC experiments  \cite{Cerri:2018ypt}, particularly with the LHCb phase II upgrade \cite{Aaij:2244311}, and from Belle II \cite{Kou:2018nap}, will further clarify the current picture. This will hopefully lead both to a clearer evidence for NP, possibly supported by other complementary set of measurements~\cite{Gherardi:2019zil,Fuentes-Martin:2019mun,Alasfar:2020mne,Crivellin:2020oup}, and to an improved understanding of the QCD dynamics of charm contributions.

\noindent \textit{Acknowledgements.} The work of M.F. is supported by project C3b of the DFG-funded Collaborative Research Center TRR 257, ``Particle Physics Phenomenology after the Higgs Discovery''. The work of M.V. is supported by the NSF Grant No.~PHY-1915005. This work was supported by the Italian Ministry of Research (MIUR) under grant PRIN 20172LNEEZ.

\bibliography{hepbiblio}

\end{document}